\documentclass[shortnote,twocolumn]{jpsj3}
\usepackage{graphicx}
\usepackage{color}
\usepackage{ulem}

\def\runauthor{Y. Nishikubo {\it et al.}}

\title{Superconductivity in the Honeycomb-Lattice Pnictide SrPtAs}

\author{Yoshihiro \textsc{Nishikubo}$^{1,2}$, Kazutaka \textsc{Kudo}$^{1,2}$\thanks{E-mail address: kudo@science.okayama-u.ac.jp}, \\ and Minoru \textsc{Nohara}$^{1,2}$}
\inst{$^1$Department of Physics, Faculty of Science, Okayama University, Okayama  700-8530, Japan \\ 
$^2$Transformative Research-Project on Iron Pnictides (TRIP), Japan Science and Technology Agency (JST), Chiyoda-ku, Tokyo 102-0075, Japan} 

\kword{pnictide superconductor, AlB$_2$-type structure, honeycomb lattice, SrPtAs}

\begin{document}
\maketitle

Transition-metal pnictides form a fascinating class of superconductors. 
Their representatives are iron-based materials\cite{rf:Kamihara,rf:Ishida}, which include REFeAsO (RE $=$ rare earth elements), AEFe$_2$As$_2$ (AE $=$ alkali earth elements), and AFeAs (A $=$ alkali elements). 
The highest transition temperature $T_{\rm c}$ to date has reached 56 K in Th-substituted GdFeAsO\cite{rf:CWang}. 
Pnictides without iron also exhibit superconductivity, though $T_{\rm c}$ is markedly lower than that of the iron-based superconductors. 
For example,  $T_{\rm c} = $ 3 and 0.7 K in BaNi$_2$P$_2$ \cite{rf:Mine} and BaNi$_2$As$_2$ \cite{rf:Ronning} with a ThCr$_2$Si$_2$-type structure, respectively. 
$T_{\rm c} =$ 5.2 K in SrPt$_2$As$_2$\cite{rf:Kudo} with a CaBe$_2$Ge$_2$-type structure. 
These superconducting pnictides share a common crystal structure with a square lattice of transition-metal elements.

A honeycomb lattice is also an attractive playground for superconductivity. 
MgB$_2$ ($T_{\rm c} =$ 39 K)\cite{rf:Nagamatsu} with an AlB$_2$-type structure, CaAlSi ($T_{\rm c} = 5.68 - 7.7$ K)\cite{rf:Imai,rf:Kuroiwa1} with an AlB$_2$ derivative structure, and CaC$_6$ ($T_{\rm c} =$ 11.5 K)\cite{rf:Weller,rf:Emery} with an  intercalated-graphite structure exhibit superconductivity with a relatively high $T_{\rm c}$ owing to a strong electron-phonon coupling\cite{rf:Kuroiwa1,rf:Walti,rf:Kuroiwa2}, soft phonons\cite{rf:Baron,rf:Kuroiwa3}, multibands/multigaps\cite{rf:Souma}, and interlayer bands\cite{rf:Okazaki}.

In this paper, we report superconductivity in SrPtAs with $T_{\rm c} =$ 2.4 K. 
To our knowledge, this is the first superconducting pnictide with a honeycomb lattice structure. 
SrPtAs crystallizes in a hexagonal KZnAs-type structure with the space group $P6_3/mmc$ (\#194)\cite{rf:Wenski}. 
This structure is derived from the binary AlB$_2$-type structure with the space group $P6/mmm$ (\#191). 
The schematic views of AlB$_2$ and SrPtAs are shown in Fig. 1. 
In AlB$_2$, boron atoms form honeycomb layers and aluminum atoms are intercalated between them. 
In SrPtAs, the aluminum sites are occupied by strontium atoms and the boron sites are occupied by either platinum or arsenic atoms so that they alternate in the honeycomb layer as well as in the $c$-axis. 
Thus, the structure can be viewed as an ordered variant of the AlB$_2$-type structure. 
\begin{figure}[t]
\begin{center}
\includegraphics[width=1\linewidth]{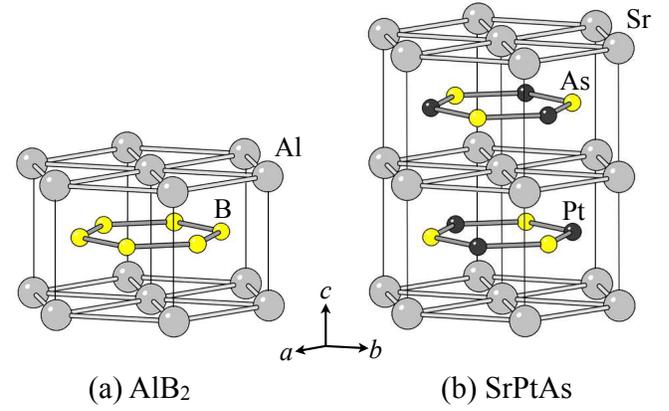}
\end{center}
\caption{(Color online) Crystal structures of (a) AlB$_2$ (space group $P6/mmm$) and (b) SrPtAs (space group $P6_3/mmc$). 
}
\label{f1}
\end{figure} 
\begin{figure}[t]
\begin{center}
\includegraphics[width=1\linewidth]{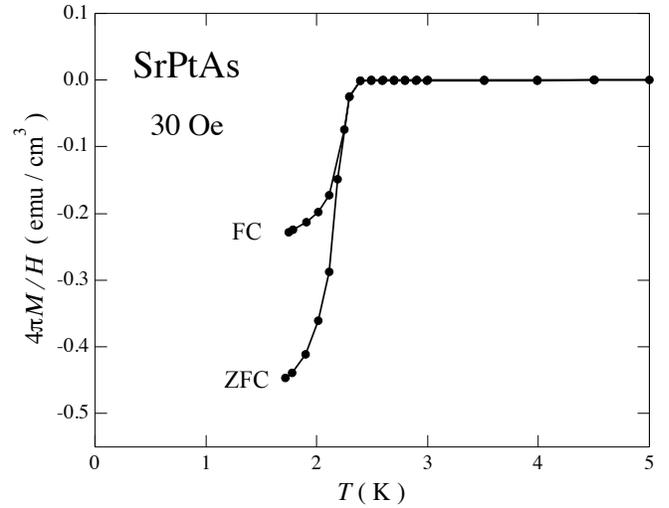}
\end{center}
\caption{Temperature dependence of magnetization divided by applied field, $M/H$, of SrPtAs at 30 Oe under zero-field-cooling (ZFC) and field-cooling (FC) conditions. 
}
\label{f2}
\end{figure}
\begin{figure}[t]
\begin{center}
\includegraphics[width=1\linewidth]{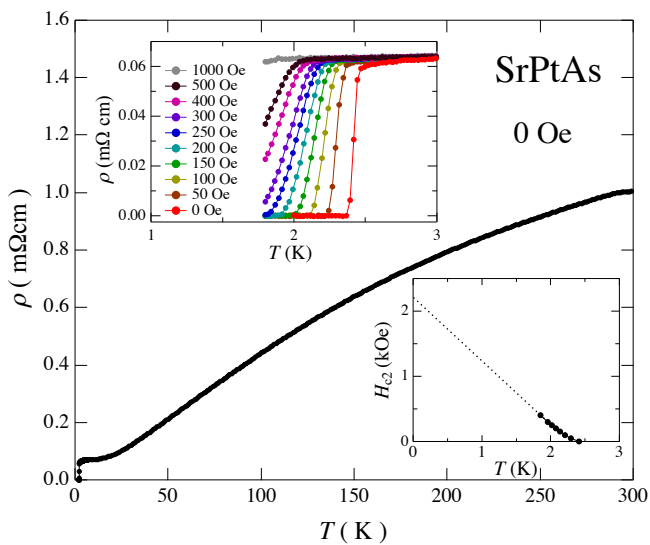}
\end{center}
\caption{(Color online) Temperature dependence of electrical resistivity $\rho$ for SrPtAs. 
The upper inset shows $\rho$ in magnetic fields. 
The lower inset shows the temperature dependence of the upper critical field $H_{\rm c2}$. 
The dotted line indicates a linear extrapolation of the data from 1.85 to 2.08 K. 
}
\label{f3}
\end{figure}

Polycrystalline samples of SrPtAs were synthesized by a solid-state reaction. 
The PtAs$_2$ precursor was first synthesized by heating Pt powder and As grains at 700 $^\circ$C in an evacuated quartz tube. 
Then, Sr, Pt, and PtAs$_2$ powders of stoichiometric amounts were mixed and ground. 
The resulting powder was placed in an alumina crucible and sealed in an evacuated quartz tube. 
The ampule was heated at 700 $^\circ$C for 3 h and then at 1000 $^\circ$C for 24 h. 
After furnace cooling, the sample was ground, pelletized, wrapped with Ta foil, and heated at 950 $^\circ$C for 2 h in an evacuated quartz tube. 
The products were confirmed to be a single phase of SrPtAs by powder X-ray diffraction. 
Lattice parameters were estimated to be $a =$ 4.244 \AA\ and $c =$ 8.989 \AA, consistent with the previous report\cite{rf:Wenski}. 
The magnetization $M$ was measured from 1.7 to 5 K under a magnetic field of 30 Oe with the Magnetic Property Measurement System (Quantum Design). 
The electrical resistivity $\rho$ was measured by the standard DC four-terminal method in the temperature range between 1.8 and 300 K under magnetic fields up to 1000 Oe using the Physical Property Measurement System (Quantum Design).

Figure 2 shows the temperature dependence of magnetization divided by the applied field, $M/H$, of SrPtAs at 30 Oe under zero-field-cooling and field-cooling conditions. 
$M$ exhibited a diamagnetic behavior below about 2.4 K, indicating the occurrence of superconductivity at $T_{\rm c} =$ 2.4 K. 
The shielding and flux exclusion signals at 1.7 K correspond to 45 and 23\% of perfect diamagnetism, respectively. 
The data support the appearance of bulk superconductivity at $T_{\rm c} =$ 2.4 K in SrPtAs.

Figure 3 shows the temperature dependence of $\rho$ for SrPtAs. 
$\rho$ exhibited a metallic behavior and clearly showed a superconducting transition at low temperatures. 
The upper inset of Fig. 3 shows the resistive transition in detail. 
The onset temperature determined from the 10\% rule was 2.45 K, and the 10$-$90\% transition width was about 0.06 K. 
The temperature where $\rho$ points to zero was 2.37 K.

As shown in  the upper inset of Fig. 3, $T_{\rm c}$ gradually decreased with increasing external magnetic field. 
The temperature dependence of the upper critical field  $H_{\rm c2}$ was determined from the midpoint of the resistive transition, as shown in the lower inset of Fig. 3. 
$H_{\rm c2}(T)$ exhibited an upward curvature with a visible change in the slope d$H_{\rm c2}$/d$T$ near $T_{\rm c}$. 
Thus, we tentatively estimated $H_{\rm c2}(0)$ to be 2200 Oe from a linear extrapolation of the data at a lower temperature part between 1.85 and 2.08 K. 
The Ginzburg-Landau coherence length $\xi_{\rm 0}$ was obtained as 387 \AA \ from $\xi_{\rm 0} = [\Phi_{\rm 0}/2\pi H_{\rm c2}(0)]^{1/2}$, where $\Phi_{\rm 0}$ is the magnetic flux quantum. 
Here, it is worth noting that the upward curvature of $H_{\rm c2}(T)$ suggests a multiband/multigap superconductivity in the present compound. 
A similar behavior has been reported in MgB$_2$\cite{rf:Gurevich1,rf:Gurevich2}, LaFeAsO$_{0.89}$F$_{0.11}$\cite{rf:Hunte}, and Ba(Fe$_{1-x}$Co$_x$)$_2$As$_2$\cite{rf:Ni}.

In summary, we observed superconductivity at $T_{\rm c} =$ 2.4 K in the honeycomb-lattice pnictide SrPtAs.
To our knowledge, this is the first pnictide superconductor with a honeycomb lattice. 
Our result shed light on a novel route to develop pnictide superconductors. 

\section*{Acknowledgement}
Part of this work was performed at the Advanced Science Research Center, Okayama University.

\end{document}